\documentclass[pre,noshowpacs,showkeys,preprint,amsmath,amsfonts]{revtex4}
\usepackage{graphicx}
\usepackage{epstopdf}
\usepackage{epsfig}
\usepackage{textcomp}
\usepackage[utf8]{inputenc}
\usepackage[english]{babel}
\usepackage[usenames, dvipsnames]{color}
\usepackage[colorlinks=true]{hyperref}

\hypersetup{colorlinks=true,citecolor=blue,linkcolor=blue,urlcolor=blue}

\usepackage{color}
\usepackage{soul}

\begin{document}

\title{Effect of discrete breathers on the specific heat of a nonlinear chain}

\author{Mohit Singh$^1$}
\email{mohitsingh1997@gmail.com}
\author{Alina~Y.~Morkina$^2$}
\email{alinamorkina@yandex.ru}
\author{Elena~A.~Korznikova$^{2,3}$}
\email{elena.a.korznikova@gmail.com}
\author{Volodymyr~I.~Dubinko$^4$}
\email{vlad@quantumgravityresearch.org}
\author{Dmitry~A.~Terentiev$^5$}
\email{dterenty@sckcen.be}
\author{Daxing Xiong$^6$}
\email{phyxiongdx@fzu.edu.cn}
\author{Oleg~B.~Naimark$^7$}
\email{naimark@icmm.ru}
\author{Vakhid~A.~Gani$^{8,9}$}
\email{vagani@mephi.ru}
\author{Sergey~V.~Dmitriev$^{3,10}$}
\email{dmitriev.sergey.v@gmail.com}

\affiliation{
$^1$Indian Institute of Technology Kharagpur, Kharagpur 721302, India\\
$^2$Ufa State Aviation Technical University, Ufa 450008, Russia\\
$^3$Institute for Molecule and Crystal Physics, Ufa Federal Research Centre of the Russian Academy of Sciences, Ufa 450075, Russia\\
$^4$NSC Kharkov Institute of Physics and Technology, Kharkov 61108, Ukraine\\
$^5$SCK CEN, Nuclear Materials Science Institute, Boeretang 200, Mol, 2400, Belgium\\
$^6$Department of Physics, Fuzhou University, Fuzhou 350108, Fujian, China\\
$^7$Institute of Continuous Media Mechanics, Ural Branch of RAS, Perm 614013, Russia\\
$^8$National Research Nuclear University MEPhI (Moscow Engineering Physics Institute), Moscow 115409, Russia\\
$^9$Institute for Theoretical and Experimental Physics of National Research Centre ``Kurchatov Institute'', Moscow 117218, Russia\\
$^{10}$Institute of Mathematics with Computing Centre, UFRC, Russian Academy of Sciences, Ufa 450008, Russia
}


\begin{abstract}
A nonlinear chain with six-order polynomial on-site potential is used to analyze the evolution of the total to kinetic energy ratio during development of modulational instability of extended nonlinear vibrational modes. For the on-site potential of hard-type (soft-type) anharmonicity, the instability of $q =\pi$ mode ($q = 0$ mode) results in the appearance of long-living discrete breathers (DBs) that gradually radiate their energy and eventually the system approaches thermal equilibrium with spatially uniform and temporally constant temperature. In the hard-type (soft-type) anharmonicity case, the total to kinetic energy ratio is minimal (maximal) in the regime of maximal energy localization by DBs. It is concluded that DBs affect specific heat of the nonlinear chain and for the case of hard-type (soft-type) anharmonicity they reduce (increase) the specific heat.
\end{abstract}

\maketitle
	

\section{Introduction}
\label{Introduction}

Discrete breathers (DBs) or intrinsic localized modes (ILMs) are spatially localized, large-amplitude oscillations in a nonlinear defect-free lattice. DBs have been discovered three decades ago by theoreticians in one-dimensional nonlinear lattices~\cite{1,2,3} and their properties have been extensively studied, as summarized in \cite{19,20}.

There exist a number of physical systems where the existence of DBs has been proven experimentally, among them are macroscopic spring-mass chains and arrays of coupled pendula or magnets~\cite{SM,pendula,Mag}, granular crystals~\cite{Gran0,Gran1,Gran2,Gran3,Gran4,Gran5,Gran6,Gran7}, micro-mechanical cantilever arrays~\cite{Cant1,Cant2,Cant3}, electrical lattices~\cite{ElLatt1,ElLatt2,ElLatt3}, nonlinear optical devices~\cite{NonlinOptics}, Josephson junction arrays \cite{Josephson1,Josephson2}. 

Crystal lattices can also accommodate DBs~\cite{UFN} since discreteness of media and non-linearity are the two prerequisites for their existence, and interatomic interactions are indeed anharmonic. A number of successful experimental studies showed the existence of DBs in crystals by measuring the vibrational spectra. The examples include DBs found in alpha-uranium~\cite{UthermalExp,UheatCapac,Uranium1}, helium~\cite{Helium}, NaI~\cite{NaI1,NaI2}, graphite~\cite{Graphite}, and PbSe~\cite{PbSe}. The concentration of DBs in crystals under thermal equilibrium conditions is relatively low~\cite{Sievers}.

Contrary to the stable topological lattice defects, e.g.\ point defects, dislocations or grain boundaries, a direct experimental observation of DBs in crystals is challenging due to their nanometeric characteristic size and short (picosecond) lifetime. That is why computer simulation methods play an important role in helping the study of DB properties in various crystals. Earlier, the existence of DBs in strained graphene and graphane (fully hydrogenated graphene) has been confirmed with the help of {\em ab initio} simulations~\cite{AbInitgraphene,AbInitgraphane}. Such first-principle simulations impose high computational demands and at present their application is limited to two-dimensional (2D) structures supporting highly localized DBs that can be analyzed in relatively small computational cells. DBs in 3D crystals are studied by means of classical molecular dynamics (MD) methods. For the first time, this method was successfully applied to the study of gap DBs in alkali halide NaI crystal~\cite{NaIDBMD1} and this study was continued in \cite{NaIDBMD2,NaIDBMD3}. Using molecular dynamics, DBs have been found in monoatomic Morse crystals~\cite{Morse1,Morse2}, covalent crystals Si, Ge and diamond~\cite{SiGe,Diamond}, pure metals~\cite{M1,M2,M2a,M2b,M3,M4,M5,M6,M7}, ordered alloys~\cite{OA0,OA1,OA2,OA3}, carbon and hydrocarbon nanomaterials~\cite{CH1,CH2,CH3,CH4,CH5,CH6,CH7,CH8,CH9,CH10,CH11,CH12,CH13}, boron nitride~\cite{CH14}, and proteins~\cite{Protein0,Protein1,Protein2,Protein3}. Essential limitation of any MD model is the choice of the interatomic potentials which largely determines the reliability of the obtained results~\cite{CH12}.

An alternative approach to investigation of the role played by DBs is to predict how they could alter macroscopic properties of crystals depending on the ambient temperature~\cite{Manley}. Since DBs are nonlinear vibrational modes, their excitation is expected to be triggered by raising the temperature above a certain threshold value~\cite{ThermalEq0}. In several experimental works, the effect of DBs on macroscopic properties of crystals has been discussed. In particular, anomalies in thermal expansion~\cite{UthermalExp} and heat capacity~\cite{UheatCapac} of alpha-uranium were attributed to the excitation of DBs at high temperatures. At the same time, it was shown numerically that DBs are responsible for the transition from ballistic to normal thermal conductivity in a nonlinear chain~\cite{Coolig12,Coolig14}.

Identification of thermally excited DBs in lattices requires application of special procedures~\cite{NaIDBMD3,ThermalEq0,ThermalEq2,ThermalEq3,ThermalEq4,ThermalEq5,ThermalEq6,ThermalEq7}. DBs, as dynamical lattice defects, interact with phonons~\cite{DBph}, and they can be distinguished in lattices during non-equilibrium processes, e.g., by absorbing running phonon waves at the boundaries of heated lattice~\cite{Protein2,Coolig12,Coolig1,Coolig2,Coolig3,Coolig4,Coolig5,Coolig6,Coolig7,Coolig9,Coolig10,Coolig11,Coolig13}.
Here we choose an alternative approach related to modulational instability of particular delocalized vibrational modes. Such instabilities lead to energy localization in the form of long-living chaotic DBs and subsequent transition to thermal equilibrium~\cite{ElLatt3,Burlakov,Mirnov,Ullmann,Kosevich,Cretegny,Mi1,Mi2,Mi3,Mi4,Mi5}. In the course of this transition of a nonlinear chain, the local temperature and the specific heat can be calculated. Here, we demonstrate that the specific heat of the crystal containing DBs is different (smaller for the hard-type anharmonicity and larger for the soft-type anharmonicity) from the one measured under thermal equilibrium. This feature can be used as indicator of the activation of DBs during increasing the crystal’s temperature and measuring its specific heat at the same time. The specific heat and the anomaly of thermal conductivity in the presence of DBs can be linked also to the definition of the effective temperature related to the additional degrees of freedom of out-of-equilibrium systems.

Discrete breathers can be divided into two large groups according to the type of anharmonicity. DBs with a soft (hard) type of anharmonicity demonstrate a decrease (increase) in the frequency of oscillations with amplitude. Obviously, discrete breathers with a soft type of anharmonicity can exist only in crystals with a gap in the phonon spectrum; examples are NaI~\cite{NaI1,NaI2,NaIDBMD1,NaIDBMD2,NaIDBMD3}, graphane~\cite{AbInitgraphane}, strained graphene~\cite{AbInitgraphene,CH11}, ordered alloys~\cite{OA0,OA1,OA2,OA3}. DBs with the hard-type anharmonicity have been reported in pure metals~\cite{UthermalExp,UheatCapac,Uranium1,M1,M2,M2a,M2b,M3,M4,M5,M6,M7} and covalent crystals~\cite{SiGe,Diamond}. It seems important to study the effect of DBs on the macroscopic properties for both groups of crystals. In our previous work~\cite{EPJB}, the Fermi--Pasta--Ulam chain was considered, where the absence of an on-site potential made it possible to study the effect of DBs on thermal expansion and elastic constants. Nevertheless, within the framework of that model, only hard-type anharmonicity DBs could be analyzed due to the absence of a gap in the phonon spectrum. Here we construct a model with the ability to consider both types of anharmonicity, introducing a six-order polynomial on-site potential.

In this work, we discuss this methodology and provide computational assessment to evaluate the contribution of DBs in the change of the total to kinetic energy ratio, which is related to the specific heat. After description of the model and simulation procedure provided in Sec.~\ref{SimulationSetup}, the modulational instability analysis is performed in Sec.~\ref{MI} for the chains with soft- and hard-type anharmonicity. Then the simulation results on the development of modulational instability of the zone boundary mode ($q=\pi$) are presented in Sec.~\ref{ResultsHard} for the hard-type anharmonicity. The modulational instability of $\Gamma$-point mode ($q=0$) is studied in Sec.~\ref{ResultsSoft} for the soft-type anharmonicity. Properties of discrete breathers are then analyzed in Sec.~\ref{DB} to rationalize results of the performed simulations. Summary and conclusions are presented in Sec.~\ref{Conclusion}.

\section{The model and simulation setup}
\label{SimulationSetup}

\begin{figure}[t!]
\includegraphics[width=9.5cm]{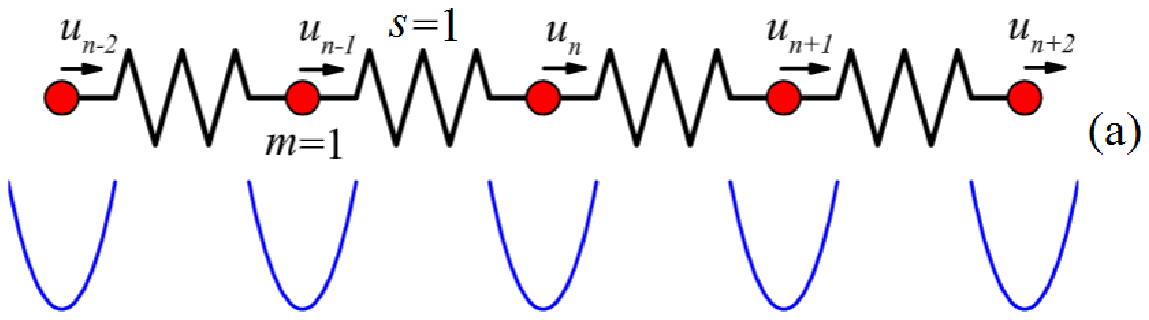}
\includegraphics[width=9.5cm]{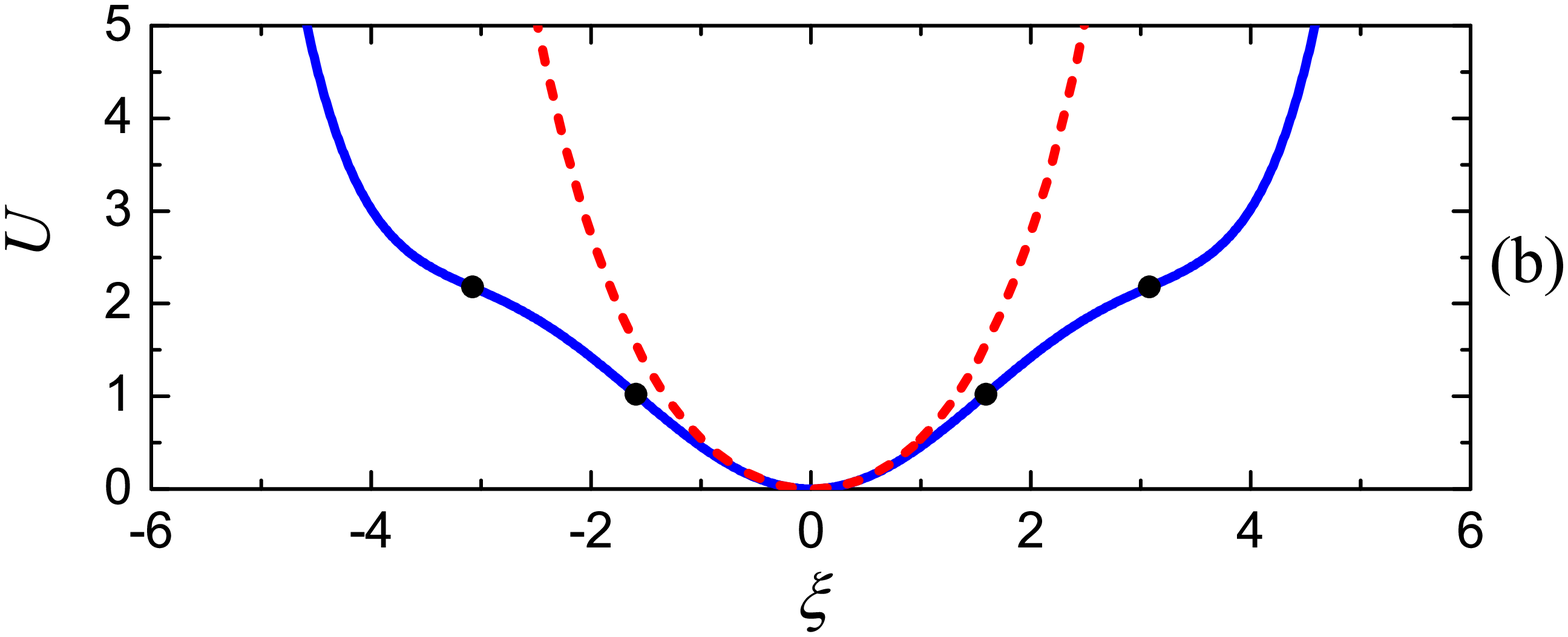}
\caption{(a) Chain of harmonically coupled, unit mass point-like particles interacting with the six-order polynomial on-site potential. (b) On-site potential Eq.~(\ref{PotentialHard}) for the cases of soft-type (blue solid line) and hard-type (red dashed line) anharmonicity. Potential parameters are $k=1/2$, $\alpha=\pm 1/24$, and $\beta=1/720$, where positive and negative $\alpha$ corresponds to the hard- and soft-type anharmonicity, respectively. The soft-type potential has inflection points at $\xi^*_i=\pm\sqrt{6\pm 6/\sqrt{3}}$, i.e., $\xi^*_{1,2}\approx \pm 1.59$ and $\xi^*_{3,4}\approx \pm 3.08$ (indicated by dots).}
\label{fig1}
\end{figure}

We consider a 1D chain of particles having mass $m$ (see Fig.~\ref{fig1}) whose Hamiltonian is defined by
\begin{equation}\label{Hamiltonian}
H = K + P = \sum_n\frac{m\dot{u}_n^2}{2} + \sum_n\left[\frac{s}{2}(u_{n+1}^{}-u_{n}^{})^2 + U(u_{n}^{})\right],
\end{equation}
where $K$ is the kinetic energy, $P$ is the potential energy, $u_n^{}$ is the displacement of the $n$th particle from its equilibrium position and $\dot{u}_n^{}$ is its velocity (overdot means derivative with respect to time $t$). The particles are harmonically coupled to their nearest neighbors by the elastic bonds with stiffness $s$. For the on-site potential, we take
\begin{equation}\label{PotentialHard}
U(\xi) = k \xi^2 + \alpha \xi^4 + \beta \xi^6,
\end{equation}
where $k$ is the coefficient in front of the harmonic term, while the coefficients $\alpha$ and $\beta$ define the contributions from the quartic and six-order terms, respectively. This model has been considered in~\cite{Kuzkin} for solving a different problem. Without loss in generality we set bond length equal to unity. By a proper choice of the units of time and energy we can set $m=1$ and $s=1$, respectively.

As mentioned in the Introduction, it is important to analyze the effect of DBs on macroscopic properties of the lattices for the cases of soft- and hard-type anharmonicity. In the framework of the present work, this can be done by choosing the following parameters for the on-site potential \eqref{PotentialHard}: $k=1/2$, $\alpha=\pm 1/24$, and $\beta=1/720$ [see Fig.~\ref{fig1}(b)]. Note that for $\alpha>0$ we have the on-site potential with the hard-type anharmonicity and for $\alpha<0$ the on-site potential features the soft-type anharmonicity for not very large vibration amplitudes. The value of the parameter $\beta$ is such that the on-site potential for the soft-type anharmonicity is the Taylor series expansion of the Frenkel--Kontorova sinusoidal potential, which will simplify the analysis of DBs in Sec.~\ref{DBsoft}. The on-site potential is shown in Fig.~\ref{fig1}(b) by the blue solid and red dashed lines for the cases of soft- and hard-type anharmonicity. It can be seen that in both cases the potential has single well. The soft-type potential has inflection points at $\xi^*_i=\pm\sqrt{6\pm 6/\sqrt{3}}$, i.e., $\xi^*_{1,2}\approx \pm 1.59$ and $\xi^*_{3,4}\approx \pm 3.08$ (indicated by dots). If particle displacements exceed the inflection points $\xi^*_{3,4}$ then even for negative $\alpha$ the on-site potential will become effectively hard due to the six-order term, which is introduced in order to prevent bond breaking at large vibration amplitudes.

The equations of motion that stem from Eqs.~\eqref{Hamiltonian} and \eqref{PotentialHard} are
\begin{equation}\label{EMo}
m\ddot{u}_n^{} = s\left(u_{n-1}^{}-2u_n^{}+u_{n+1}^{}\right) - 2ku_n^{} - 4\alpha u_n^3 - 6\beta u_n^5.
\end{equation}
These equations are integrated numerically using the St\"ormer method~\cite{Bakhvalov} of the sixth order with the time step $\tau=10^{-3}$. This symplectic method is very efficient in solving the Cauchy problem for a set of second-order ordinary differential equations not containing $\dot{u}_n^{}$. Symplectic integrators are popular because they allow the total energy of the system to be controlled with high precision~\cite{Symplectic1,Symplectic2,Symplectic3}. In our simulations, the total energy is conserved with a relative error not exceeding $10^{-5}$ during the entire numerical run.

In the case of small amplitude vibrations, the nonlinear terms can be neglected and the solutions of the linearized equation are the normal modes $u_n^{} \sim \exp [i (q n -\omega_q^{} t)]$ with the wave number $q$ and frequency $\omega_q^{}$. These modes obey the following dispersion relation:
\begin{equation}\label{Dispersion}
\omega_q^2 = \frac{2}{m}\left[k+s\left(1-\cos q\right)\right].
\end{equation}
The considered chain supports the small-amplitude running waves (phonons) with frequencies within the band from $\omega_{\min}^{}=1$ for $q=0$ to $\omega_{\max}^{}=\sqrt{5}\approx 2.236$ for $q=\pi$.

In the case of hard-type anharmonicity ($\alpha=1/24$), the zone-boundary mode with $q=\pi$ and the amplitude $A$,
\begin{equation}\label{ZBmode}
u_n^{} = A\sin(\pi n-\omega_{\max}^{}t),
\end{equation}
is excited in the chain of $N=1000$ particles at $t=0$. For the chain with soft-type anharmonicity ($\alpha=-1/24$) the $\Gamma$-point mode with $q=0$ and the amplitude $A$,
\begin{equation}\label{Gmode}
u_n^{} = A\sin(\omega_{\min}^{}t),
\end{equation}
is initially excited.

Note that the modes \eqref{ZBmode} and \eqref{Gmode} are lattice symmetry dictated exact solutions to the equations of motion regardless the type of interaction potential and for arbitrary amplitude. There exist other modes with short periods having the same properties~\cite{Bush1,Bush2,Bush3}. However those modes, at least for small amplitudes, typically have frequencies within the phonon band and the development of their instability does not produce long-lived chaotic DBs.

If $A$ is not too small, the modes \eqref{ZBmode} and \eqref{Gmode} are modulationally unstable. Initially the energy is evenly shared between all the particles. Development of the instability results in energy localization which can be monitored by calculating the localization parameter 
\begin{equation}\label{Localiz}
L=\frac{\sum e_n^2}{\Big(\sum e_n^{}\Big)^2},
\end{equation}
where 
\begin{equation}\label{en}
e_n^{} = \frac{m\dot{u}_n^2}{2} + \frac{s}{4}\left(u_n^{}-u_{n-1}^{}\right)^2 + \frac{s}{4}\left(u_{n+1}^{}-u_n^{}\right)^2 + U(u_n^{})
\end{equation}
is the energy of the $n$th particle.

In our study, the energies of particles $e_n^{}$ are chosen as observables, as in a number of other studies on DBs~\cite{ThermalEq4,Observables2,Observables3}. The use of such local observables is justified for the high energy regime when they are constants of motion. On the other hand, in the low energy regimes, energies of normal modes can be used as observables, as they become constants of motion~\cite{qBreathers}. In the present study, it seems that we have an intermediate case which cannot be classified as the low energy regime because, as it will be seen below, chaotic DBs emerging as a result of modulational instability, are highly localized and have relatively large amplitudes. This is why the local observables $e_n^{}$ seem to be informative in our study.

As a measure of temperature, the averaged kinetic energy per atom,
\begin{equation}\label{Ken}
\bar{K} = \frac{1}{N}\sum_n\frac{m\dot{u}_n^2}{2},
\end{equation}
will be used. In fact, the temperature of a one-dimensional lattice is $T=2\bar{K}/k_{\rm B}^{}$, where $k_{\rm B}^{}=8.617\times 10^{-5}$~eVK$^{-1}$ is the Boltzmann constant.

Heat capacity of the whole chain is defined as
\begin{equation}\label{HeatCap}
C = \lim_{\Delta T\to 0} \frac{\Delta H}{\Delta T},
\end{equation}
where $\Delta H$ is the portion of energy given to the system and $\Delta T$ is the corresponding increase of temperature. Specific heat is defined as the heat capacity per unit mass or per particle. Since periodic boundary conditions are used in this study, and thermal expansion of the chain is not allowed, we evaluate the specific heat at constant volume.

The definition \eqref{HeatCap} cannot be used in our simulations performed at constant total energy $H$. An alternative way to characterize heat capacity is to consider the total energy to kinetic energy ratio for the chain~\cite{Ebeling1,Ebeling2}. In linear chains the total energy is equally shared between the kinetic and potential energies so that $\Delta  H=2\Delta  K$ and $C=2$, while the portion of the kinetic energy can differ from 1/2 in a nonlinear chain. In this study the specific heat of the chain at constant volume is characterized by the ratio
\begin{equation}\label{HeatCapHere}
c_V^{}=\frac{\bar{H}}{\bar K},
\end{equation}
where $\bar H$ and $\bar K$ are the total energy and the kinetic energy of the chain per atom, respectively. 

In the following section, the time evolution of $c_V^{}$ will be calculated for the chain during the development of modulational instability. The values of $c_V^{}$ in the regime when energy is localized by DBs will be compared to that in thermal equilibrium.

\section{Modulational instability}
\label{MI}

Several analytical approaches have been developed for the analysis of modulational instabilities in lattices~\cite{Kosevich,KivsharPeyrard1993}. Here we follow the one proposed by Kosevich and Lepri~\cite{Kosevich}. We first derive the wavenumber of the modulation wave with the largest growth rate for the $\Gamma$-point mode and zone-boundary mode in the cases of soft-type and hard-type anharmonicity, respectively, and then compare the analytical results to the results of numerical simulations.

\subsection{Soft-type anharmonicity}
\label{MIsoft}

Let us analyze the modulational instability of the $\Gamma$-point mode \eqref{Gmode} with the amplitude $A$. Long-wavelength approximation of Eq.~\eqref{EMo} reads (the lattice spacing is equal to 1 in our model)
\begin{equation}\label{EMoSoftLW}
mu_{tt}^{} = su_{xx}^{} - 2ku-4\alpha u^3,
\end{equation}
where the quintic term is omitted assuming that $A\sim 1$ and taking into account that $\beta\ll |\alpha|$.

We look for the solution to Eq.~\eqref{EMoSoftLW} in the form
\begin{equation}\label{AnsatzSoft}
u(x,t) = A\sin(\omega_{\rm min}^{}t) + \epsilon(t)\cos(Qx),
\end{equation}
where the first term in the right-hand side is the $\Gamma$-point mode and the second term is a small perturbation $(\epsilon\ll A)$ in the form of standing wave with the wavenumber $Q$. Substituting Eq.~\eqref{AnsatzSoft} into Eq.~\eqref{EMoSoftLW} one finds the frequency of the $\Gamma$-point mode as a function of amplitude,
\begin{equation}\label{Wmin}
\omega_{\rm min}^2=\frac{2k+3\alpha A^2}{m},
\end{equation}
and the linearized equation for the evolution of the perturbative term,
\begin{equation}\label{PerturbSoft}
m\ddot{\epsilon}+[sQ^2+2k+12\alpha A^2\sin^2(\omega_{\rm min}^{} t)]\epsilon=0.
\end{equation}
This result assumes the weakness of nonresonant interaction between the mode with fundamental frequency and its third harmonic, which allows to substitute $\sin^3(\omega_{\rm min}^{}t)$ with $(3/4)\sin(\omega_{\rm min}^{}t)$.

The solution of Eq.~\eqref{PerturbSoft} is taken in the form~\cite{Kosevich}
\begin{equation}\label{SFormSoft}
\epsilon=[a\cos(\omega_{\rm min}^{} t)+b\sin(\omega_{\rm min}^{} t)]e^{pt},
\end{equation}
where $p$ is the instability growth rate. Substitution of Eq.~\eqref{SFormSoft} into Eq.~\eqref{PerturbSoft} yields 
\begin{equation}\label{SolSoft}
m^2p^4+2m(sQ^2+4k+9\alpha A^2)p^2+sQ^2(sQ^2+6\alpha A^2)=0,
\end{equation}
where again the assumption of weak interaction of the fundamental harmonic with the third harmonic was used. Physically meaningful root of Eq.~\eqref{SolSoft} for the parameters used in this study is 
\begin{equation}\label{RootSoft}
p=\sqrt{- 2+\frac{3}{8}A^2  - Q^2
+\sqrt{4 - \frac{3}{2}A^2 + \frac{9}{64}A^4 + \Big(4-\frac{1}{2}A^2\Big)Q^2}}.
\end{equation}
Condition $dp/dQ=0$ gives the wavenumber corresponding to the maximal growth rate,
\begin{equation}\label{QmaxSoft}
Q_{\max}^{}=\frac{A}{\sqrt{8}}\sqrt{ \frac{8-5A^4/4}{8-A^2}},
\end{equation}
which for small $A$ can be simplified to
\begin{equation}\label{QmaxSoftSimple}
Q_{\max}^{}=\frac{A}{\sqrt 8}.
\end{equation}
Corresponding maximal growth rate for small $A$ is
\begin{equation}\label{pmaxSoftSimple}
p_{\max}^{}=\frac{A^2}{16}.
\end{equation}

\subsection{Hard-type anharmonicity}
\label{MIhard}

Following the work~\cite{Kosevich}, for the analysis of instability of the zone-boundary mode we introduce new variable $f_n^{}=(-1)^nu_n^{}$ and rewrite Eq.~\eqref{EMo} in the form
\begin{equation}\label{EMof}
m\ddot{f}_n^{}=
-s(f_{n-1}^{}-2f_n^{}+f_{n+1}^{})-(4s+2k)f_n^{} -4\alpha f_n^3 - 6\beta f_n^5.
\end{equation}
Assuming that $f_n^{}$ varies slowly with $n$ we substitute the last equation with the long-wave approximation
\begin{equation}\label{EMofLW}
mf_{tt}^{}=
-sf_{xx}^{}-(4s+2k)f-4\alpha f^3,
\end{equation}
where the quintic term is omitted since $\beta\ll \alpha$.

The following analysis is very similar to the case of the soft-type nonlinearity. Looking for the solution to Eq.~\eqref{EMofLW} in the form
\begin{equation}\label{AnsatzHard}
f(x,t)=A\sin(\omega_{\rm max}^{}t)+\epsilon(t)\cos(Qx),
\end{equation}
one finds the frequency of the zone boundary mode as a function of amplitude,
\begin{equation}\label{Wmin1}
\omega_{\rm max}^2=\frac{4s+2k+3\alpha A^2}{m},
\end{equation}
and the linearized equation for the evolution of the modulation wave,
\begin{equation}\label{PerturbHard}
m\ddot{\epsilon}+[-sQ^2+4s+2k+12\alpha A^2\sin^2(\omega_{\rm max}^{} t)]\epsilon=0.
\end{equation}
Substitution of Eq.~\eqref{SFormSoft} into Eq.~\eqref{PerturbHard} yields 
\begin{equation}\label{SolHard}
m^2p^4+2m(-sQ^2+8s+4k+9\alpha A^2)p^2+sQ^2(sQ^2-6\alpha A^2)=0.
\end{equation}
Physically meaningful root of Eq.~\eqref{SolHard} for the parameters used in this study is 
\begin{equation}\label{RootHard}
p=\sqrt{Q^2-\frac{3}{8}A^2 - 10
+\sqrt{100 + \frac{15}{2}A^2 + \frac{9}{64}A^4 - \Big(20+\frac{1}{2}A^2\Big)Q^2}}.
\end{equation}
The wavenumber corresponding to the maximal growth rate is
\begin{equation}\label{QmaxHard}
Q_{\max}^{}=\sqrt{5}A \sqrt{\frac{1+A^2/32}{40+A^2}},
\end{equation}
which for small $A$ simplifies to
\begin{equation}\label{QmaxHardSimple}
Q_{\max}^{}=\frac{A}{\sqrt 8}.
\end{equation}
Corresponding maximal growth rate for small $A$ is
\begin{equation}\label{pmaxHardSimple}
p_{\max}^{}=\frac{A^2}{36}.
\end{equation}

Interestingly,  Eqs.~\eqref{QmaxSoftSimple} and \eqref{QmaxHardSimple} coincide for $\Gamma$-point and zone-boundary modes.

\subsection{Numerical results}
\label{MInumer}

After excitation of a $\Gamma$-point or zone-boundary mode with the amplitude $A$ we calculate energy of the particles, $e_n^{}$, according to Eq.~\eqref{en}. Initially, all particles have equal energy, but situation changes with the development of instability. We find the particles with the minimal and maximal energies, $e_{\min}^{}$, $e_{\max}^{}$, and at the time when $(e_{\max}^{}-e_{\min}^{})/(e_{\max}^{}+e_{\min}^{})>0.01$ the Fourier transform of $e_n^{}$ is performed in order to find the wavenumber of the modulation wave, $Q$.

\begin{figure}[t!]
\includegraphics[width=9.0cm]{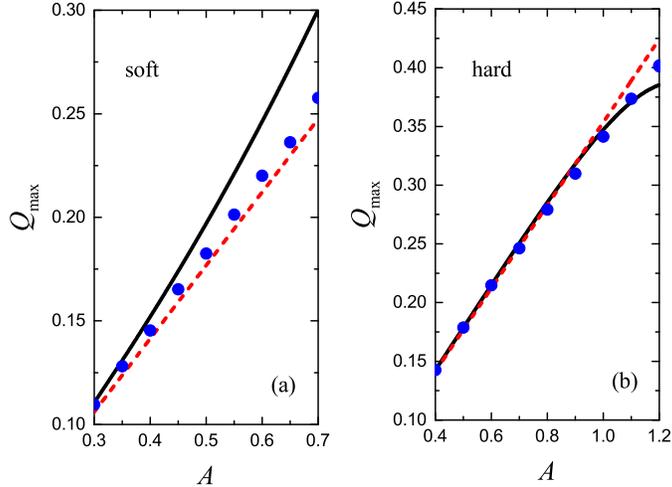}
\caption{Wavenumber of the modulation wave with the highest grows rate as a function of amplitude for (a) $\Gamma$-point mode and (b) zone-boundary mode in the cases of soft-type and hard-type anharmonicity, respectively. Numerical data are presented by dots. Black solid lines correspond to Eqs.~\eqref{QmaxSoft} and \eqref{QmaxHard}, red dashed lines correspond to Eqs.~\eqref{QmaxSoftSimple} and \eqref{QmaxHardSimple}, in (a) and (b), respectively.}
\label{figNew}
\end{figure}

In Fig.~\ref{figNew}, the numerical result is presented by dots, showing the wavenumber of the modulation wave as a function of mode amplitude for (a) $\Gamma$-point and (b) zone-boundary mode, in the cases of the soft- and hard-type anharmonicity, respectively.
Black solid lines in (a) and (b) are plotted with the use of Eqs.~\eqref{QmaxSoft} and \eqref{QmaxHard}, respectively. Red dashed lines in (a) and (b) correspond to Eqs.~\eqref{QmaxSoftSimple} and \eqref{QmaxHardSimple}, respectively, and, as it has been mentioned, they actually coincide. As expected, the analytical expressions fit the numerical results very well for the small amplitudes. More complicated expressions \eqref{QmaxSoft} and \eqref{QmaxHard} give a better result for small amplitudes, but for increasing amplitudes they deviate from the numerical results faster than the simplified expressions \eqref{QmaxSoftSimple} and \eqref{QmaxHardSimple}.  

In the following, for the soft-type (hard-type) anharmonicity the range of amplitudes from 0.35 to 0.6 (from 0.8 to 1.2) will be analyzed. Even in these ranges of amplitudes the analytical estimation provides adequate results.

\section{Chaotic discrete breathers. Hard-type anharmonicity}
\label{ResultsHard}

We take $\alpha=1/24$ in the on-site potential \eqref{PotentialHard} and excite in the chain the zone-boundary mode \eqref{ZBmode} considering various amplitudes $A$. While integrating the equations of motion \eqref{EMo}, we monitor the change in the localization parameter $L$, Eq.~\eqref{Localiz}, and specific heat $c_V^{}$, Eq.~\eqref{HeatCapHere}.

\begin{figure}[t!]
\includegraphics[width=8.0cm]{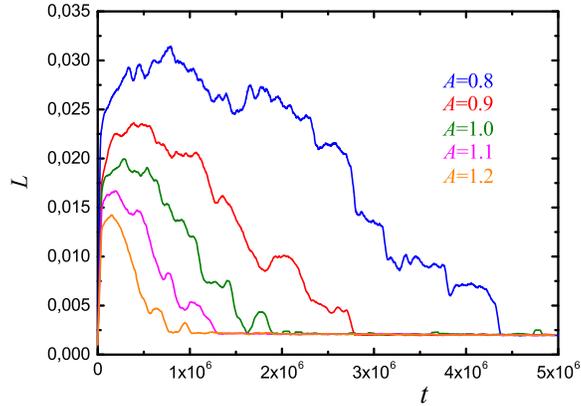}
\caption{Localization parameter vs time for various amplitudes $A$ of the initially excited zone-boundary mode ($\alpha=1/24$, hard-type anharmonicity). For all cases at $t=0$, $L=1/N=10^{-3}$. Modulational instability results in increase of $L$ due to energy localization on DBs. Then $L$ decreases because DBs gradually radiate energy and eventually the system reaches thermal equilibrium with a small value of $L$.}
\label{fig2}
\end{figure}

Localization parameter as a function of time is presented in Fig.~\ref{fig2} for various values of $A$.
At $t=0$, all curves start from the minimal possible value of the localization parameter that is $L=1/N=10^{-3}$. The development of modulational instability results in energy localization due to the formation of DBs and this leads to an increase in the localization parameter. DBs slowly radiate their energy and thus, the localization parameter gradually decreases and in the end, when the system reaches the state of thermal equilibrium, $L$ oscillates near the small value of $2\times 10^{-3}$.

\begin{figure}[t!]
\includegraphics[width=8.5cm]{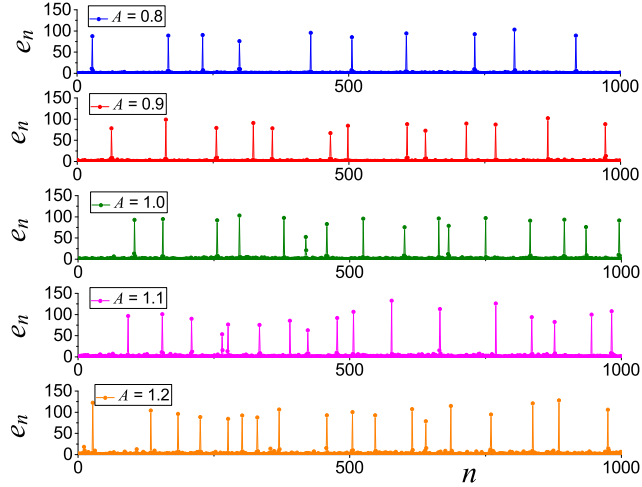}
\caption{Total energies of particles in the chain at the time when the localization parameter reaches its maximum for various amplitudes of the initially excited zone-boundary mode, as indicated for each panel. Results for $\alpha=1/24$ (hard-type anharmonicity).}
\label{fig3}
\end{figure}

Distribution of energy over the chain at the time when the localization parameter is maximal is shown in Fig.~\ref{fig3} for various mode amplitudes from $A=0.8$ to $A=1.2$.
One can see the sets of highly localized DBs. Of course, the model considered here does not support compact DBs described by Eleftheriou et al.\ for the anharmonic chain~\cite{CompactDB}, or DBs with superexponential tails explored by Dey et al.~\cite{CompactDB1}, but supports common DBs with exponentially localized tails. The tail solution can be obtained by substituting $u_n^{}(t)=d^n\sin \omega t$ into linearized equation of motion \eqref{EMo}. The resulting characteristic equation has two roots, $d_1=1/d_2=(-\kappa+\sqrt{\kappa^2-4})/2$, where $\kappa=(m\omega^2-2s-2k)/s$. The solution is valid for $u_n^{}\ll 1$. 

\begin{figure}[t!]
\includegraphics[width=8.5cm]{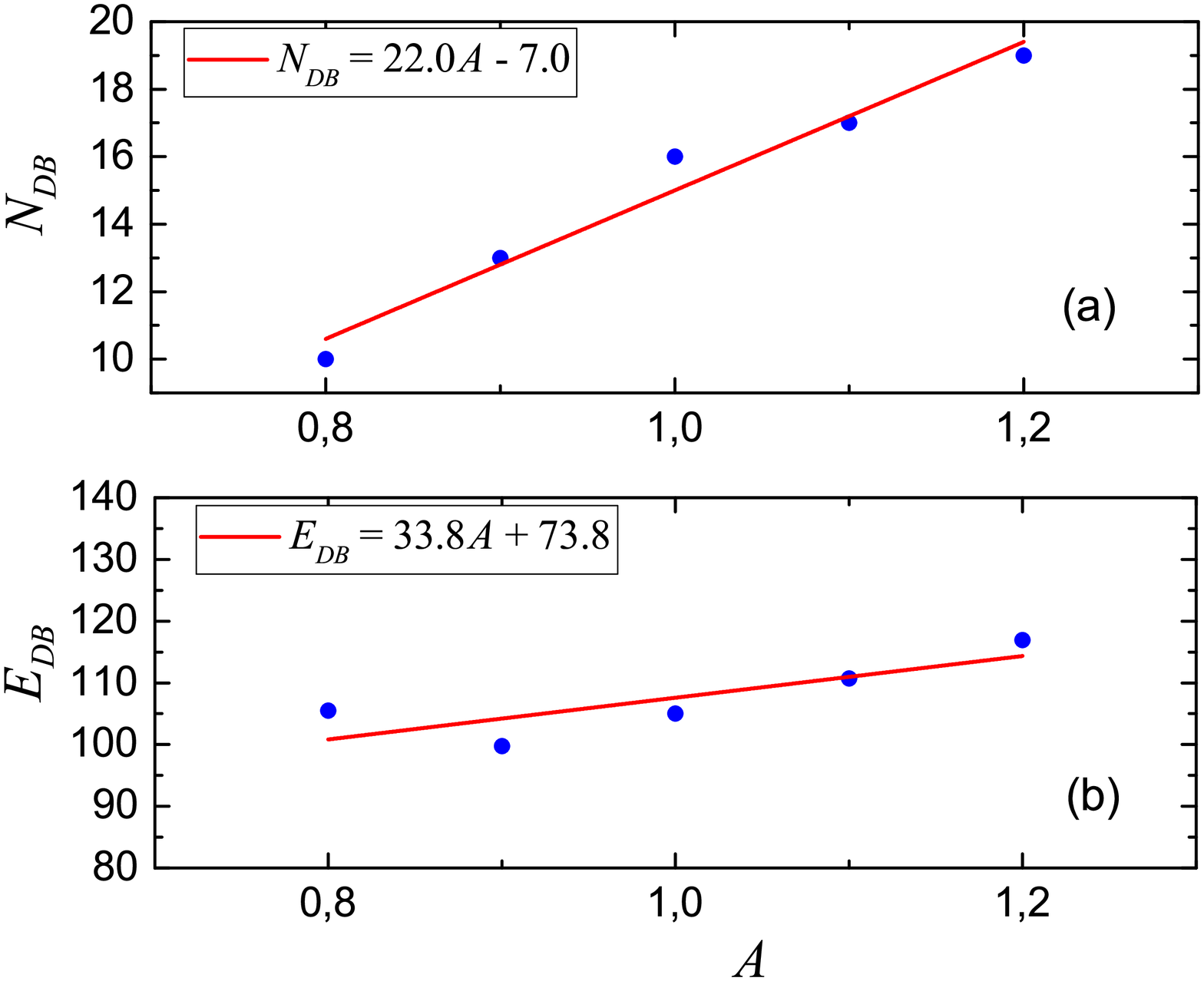}
\caption{(a) Number of discrete breathers and (b) average energy of discrete breathers at the time when localization parameter is maximal, as functions of the zone-boundary mode amplitude.}
\label{fig4}
\end{figure}

In Fig.~\ref{fig4}, we plot (a) the number of DBs and (b) the average energy of DBs as functions of the zone-boundary mode amplitude at the time when $L$ is maximal.
When calculating the number of DBs only the particles with $e_n^{}>10$ were taken into consideration. This cut-off value corresponds to about 10\% of the averaged DB energy. Doubling or even tripling the cut-off energy has very little effect on the result. It can be seen that the data shown in Fig.~\ref{fig4} has some noise. The noise could be reduced by averaging over realizations, but we did not do that in this study. It follows from the plots in Fig.~\ref{fig4} that $N_{\rm DB}^{}$ increases linearly with $A$ nearly two times within the studied range of amplitudes, while $E_{\rm DB}^{}$ increases with $A$ very slowly. 

\begin{figure}[t!]
\includegraphics[width=8.0cm]{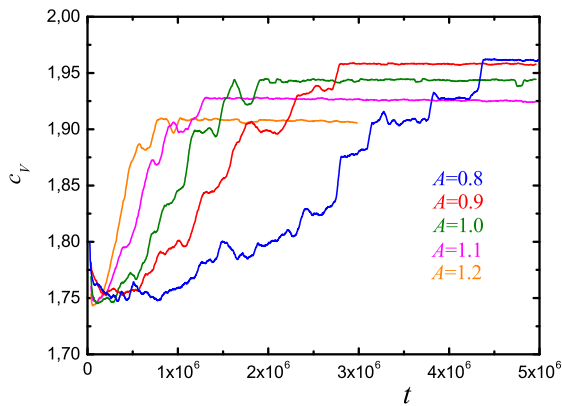}
\caption{Specific heat as a function of time for various amplitudes of the initially excited zone-boundary mode. Specific heat is minimal when DBs are in the system and it increases while system approaches thermal equilibrium. Results for $\alpha=1/24$ (hard-type anharmonicity).}
\label{fig5}
\end{figure}

Our main result is shown in Fig.~\ref{fig5}, where the time-dependence of specific heat is plotted for various mode amplitudes.
From comparison of Fig.~\ref{fig2} and Fig.~\ref{fig5}, it can be seen that the specific heat is minimal when the localization parameter is maximal. During the transition to thermal equilibrium, the specific heat increases. From this, we conclude that the DBs reduce the specific heat of the chain with hard-type anharmonicity. 

Thus, we have two distinct regimes: the regime with high localization parameter $L$, when chaotic DBs share almost all the energy of the system, and the regime with small localization parameter, when almost all the energy belongs to phonons. In the phonon regime, as follows from Fig.~\ref{fig5}, $c_V^{}$ oscillates near a constant value which decreases with increase in the total energy of the system, the latter is higher for larger $A$. In the regime of chaotic DBs the contribution to $c_V^{}$ from phonons is small and in the regime of thermal equilibrium, the contribution from DBs is small. 


\section{Chaotic discrete breathers. Soft-type anharmonicity}
\label{ResultsSoft}

Now, we take $\alpha=-1/24$ in the on-site potential \eqref{PotentialHard} and excite in the chain the $\Gamma$-point mode \eqref{Gmode} considering various amplitudes $A$. Again, while integrating the equations of motion \eqref{EMo}, we monitor the change in the localization parameter $L$, Eq.~\eqref{Localiz}, and specific heat $c_V^{}$, Eq.~\eqref{HeatCapHere}.

\begin{figure}[t!]
\includegraphics[width=8.0cm]{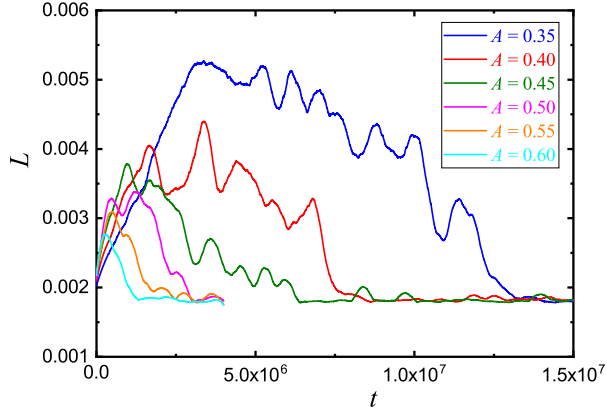}
\caption{The same as in Fig.~\ref{fig2} but for the soft-type anharmonicity ($\alpha=-1/24$).}
\label{fig6}
\end{figure}

Localization parameter as a function of time is presented in Fig.~\ref{fig6} for various values of $A$.
Similar to the case of the hard-type anharmonicity, at $t=0$, all curves start from the minimal possible value of the localization parameter that is $L=1/N=10^{-3}$. The development of modulational instability results in energy localization due to the formation of DBs, and this leads to an increase in the localization parameter. DBs slowly radiate their energy, and thus the localization parameter gradually decreases and in the end, when the system reaches the state of thermal equilibrium, $L$ oscillates near the small value of $2\times 10^{-3}$.

\begin{figure}[t!]
\includegraphics[width=9.25cm]{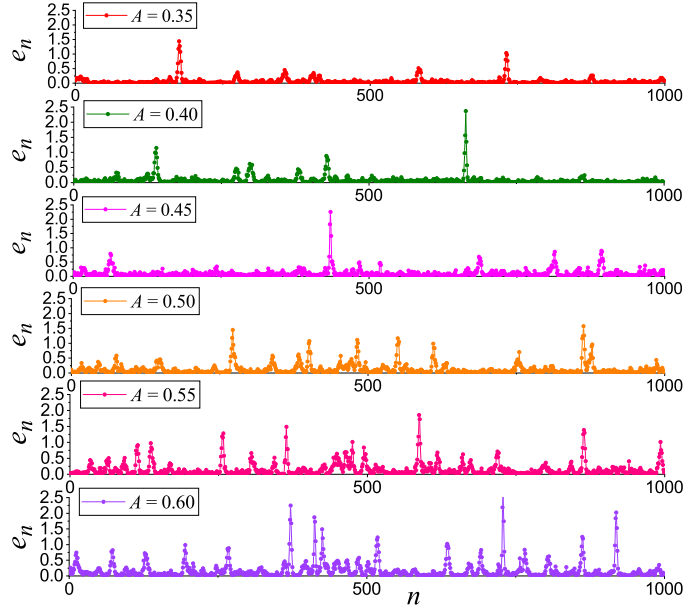}
\caption{Total energies of particles in the chain at the time when the localization parameter reaches its maximum for various amplitudes of the initially excited $\Gamma$-point mode, as indicated for each panel. Results for $\alpha=-1/24$ (soft-type anharmonicity).}
\label{fig7}
\end{figure}

\begin{figure}[t!]
\includegraphics[width=8.5cm]{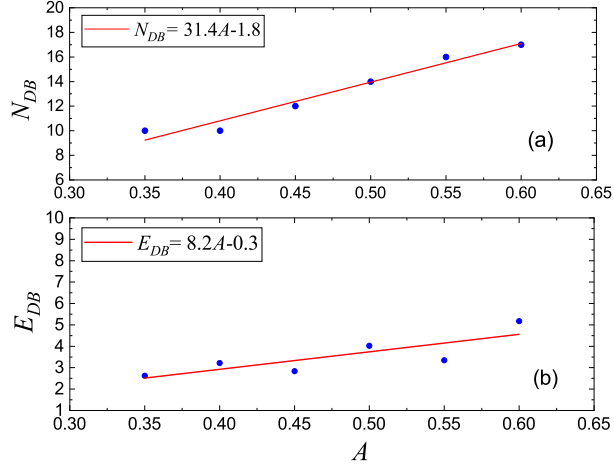}
\caption{(a) Number of discrete breathers and (b) Average energy of discrete breathers at the time when localization parameter is maximal, as the functions of the $\Gamma$-point mode amplitude.}
\label{fig8}
\end{figure}

\begin{figure}[t!]
\includegraphics[width=8.5cm]{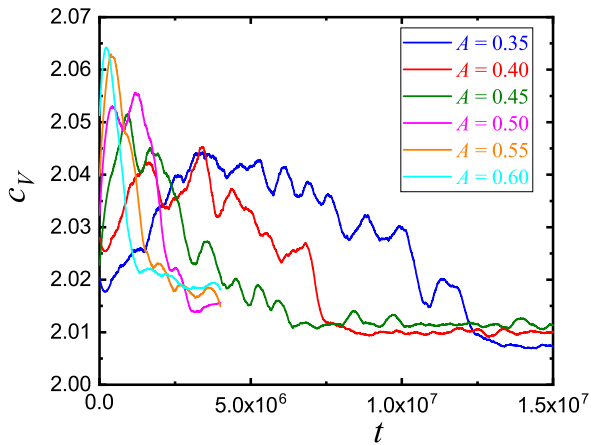}
\caption{The same as in Fig.~\ref{fig5}, but for the soft-type anharmonicity ($\alpha=-1/24$). Contrary to the case of the hard-type anharmonicity, here specific heat is maximal when DBs are in the system.}
\label{fig9}
\end{figure}

Distribution of energy over the chain at the time when localization parameter is maximal is shown in Fig.~\ref{fig7} for various mode amplitudes from $A=0.35$ to $A=0.6$.
One can see the sets of DBs localized on a few particles (they are not as sharply localized as in the case of hard-type anharmonicity). In Fig.~\ref{fig8}, we plot (a) the number of DBs and (b) the average energy of DBs as functions of the $\Gamma$-point mode amplitude at the time when $L$ is maximal.
In this case the cut-off energy for calculation of the number of DBs is set to $e_n^{}<0.3$, which is about 10\% of the averaged DB energy. It follows from the plots that $N_{\rm DB}^{}$ increases linearly with $A$ nearly two times within the studied range of amplitudes, while $E_{\rm DB}^{}$ increases with $A$ more slowly. Our main result is shown in Fig.~\ref{fig9}, where the time-dependence of the specific heat is plotted for various mode amplitudes.
From comparison of Fig.~\ref{fig6} and Fig.~\ref{fig9}, it can be seen that the specific heat is maximal when the localization parameter is maximal. During the transition to thermal equilibrium, the specific heat decreases. From this, we conclude that the DBs increase the specific heat of the chain with soft-type anharmonicity. 

In contrast to the case of the hard-type anharmonicity, here, in the phonon regime, $c_V^{}$ oscillates near a constant value which increases with increase in the total energy of the system.

\section{Properties of discrete breathers}
\label{DB}

Approximate solutions for DBs in the chain \eqref{EMo} have been derived in \cite{Kuzkin}. However, the solution reported for the hard-type anharmonicity cannot be used here because it is valid only for relatively wide DBs but in our simulations very sharp DBs are formed as a result of modulational instability, see Fig.~\ref{fig3}. In the following we will give another approximate solution, which is valid for very sharp DBs in the case of hard-type anharmonicity. On the other hand, in the case of soft-type anharmonicity the emerging DBs are not very sharp, see Fig.~\ref{fig7}, and the solution reported in \cite{Kuzkin} gives a reasonable accuracy. Below we will reproduce that solution for the convenience of the reader.

\subsection{Hard-type anharmonicity}
\label{DBhard}

For a very sharp DB localized on $n$th particle, we assume that $u_{n-1}^{} = u_{n+1}^{} = 0$. Hence, Eq.~\eqref{EMo} becomes
\begin{equation}\label{NewEMo}
    \ddot{u}_n^{} + a_1^{}u_n^{} + a_3^{}u_n^3 + a_5^{}u_n^5 = 0,
\end{equation}
where
\begin{gather}
    a_1^{} = \frac{2(s+k)}{m},\hspace{0.2cm} a_3^{} = \frac{4\alpha}{m} \hspace{0.2cm}{\rm and}\hspace{0.2cm} a_5^{} = \frac{6\beta}{m}. \nonumber
\end{gather}
The exact periodic solution to this equation has been reported, e.g., in \cite{CubicQuintic} in the form
\begin{equation}\label{EOMSolution}
u_n^{}(t) = \frac{A_{\rm DB}^{} {\rm cn}(P,M)}{\sqrt{{\rm cn^2}(P, M) + \sqrt{\frac{6q_{1}^{}}{q_{2}^{}}}{\rm sn^2}(P, M){\rm dn^2}(P, M)}},
\end{equation}
where \rm{cn}, \rm{sn} and \rm{dn} are the Jacobi elliptic functions, $A_{\rm DB}^{}$ is the DB amplitude and
\begin{gather}\label{parameters}
    q_1^{} = a_1^{} + a_3^{}A_{\rm DB}^2 + a_5^{}A_{\rm DB}^4, \\
    q_2^{} = 6a_1^{} + 3a_3^{}A_{\rm DB}^2 + 2a_5^{}A_{\rm DB}^4, \\
    q_3^{} = 4a_1^{} + 3a_3^{}A_{\rm DB}^2 + 2a_5^{}A_{\rm DB}^4, \\
    P = \Big(\frac{q_1^{}q_2^{}}{6}\Big)^{\frac{1}{4}}t,\\
    M = \frac{1}{2} - \frac{q_3^{}}{4}\sqrt{\frac{3}{2q_1^{}q_2^{}}}.
\end{gather}

\begin{figure}[t!]
\includegraphics[width=8.0cm]{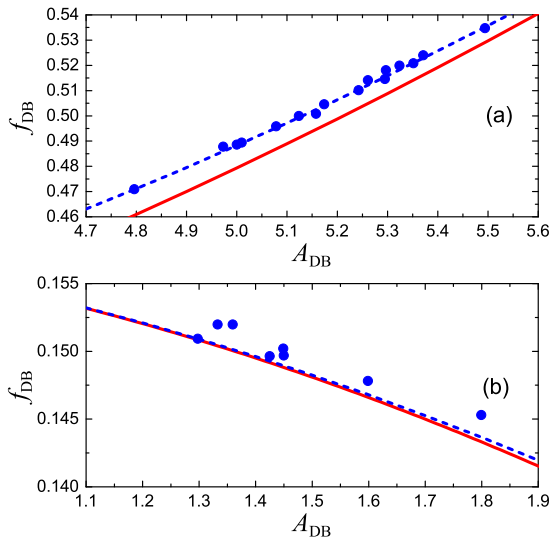}
\caption{The DB Frequency $f_{\rm{DB}}^{}$ as a function of the DB amplitude $A_{\rm{DB}}^{}$, for (a) hard-type anharmonicity and (b) soft-type anharmonicity. The red solid line in (a) is for the analytical solution \eqref{EOMSolution}, while the red solid line in (b) is for the solution \eqref{BreatherSoln}. Blue dashed lines in (a) and (b) show the numerically exact solutions for on-site DBs. Scattered dots in (a) and (b) are for the chaotic breathers emerged as a result of the modulational instability of the zone-boundary mode with the amplitude $A=1.0$ and the $\Gamma$-point mode with $A=0.6$, respectively.}
\label{fig:fig10}
\end{figure}

The red solid line in Fig.~\ref{fig:fig10}(a) shows the relation between the DB frequency ($f_{\rm{DB}}^{}$) and the DB amplitude ($A_{\rm{DB}}^{}$) obtained from the analytical solution \eqref{EOMSolution} of Eq.~\eqref{NewEMo}.
Numerically exact result for the on-site DB obtained starting from the anticontinuous limit as described in the work~\cite{Machine} is shown by the blue dashed line. Here we also plot the numerical results (scattered dots) for the chaotic DBs emerged in the simulations of the modulational instability of the zone-boundary mode with $A=1.0$, see Fig.~\ref{fig3}. It can be seen that the results for the chaotic DBs are very close to the solution obtained with the machine accuracy and the analytical solution \eqref{EOMSolution} underestimates DB frequency by approximately 5\%. This difference between the numerical and analytical results can be attributed to the assumption, that the DB is localized on single particle, used in calculating the analytical solution. This assumption is responsible for underestimation of the rigidity of the actual breather. As the DB amplitude is increased, the degree of DB localization also increases and thus, the simulation results become closer to the analytical solution at higher amplitudes.

\subsection{Soft-type anharmonicity}
\label{DBsoft}

In the case of soft-type anharmonicity, for not very large displacements, Eq.~(\ref{EMo}) can be approximated by the Frenkel-Kontorova model \cite{BraunKivshar}
\begin{equation}\label{FKeq}
    m\ddot{u}_n^{} = s(u_{n-1}^{} - 2u_n^{} + u_{n+1}^{}) - \sin(u_n^{}),
\end{equation}
which reduces to the sine-Gordon equation in the continuum limit ($s\to\infty$),
\begin{equation}\label{SGeq}
    u_{tt}^{} - u_{xx}^{} + \sin u = 0.
\end{equation}
Then the well-known moving breather solution of Eq.~\eqref{SGeq} can be written in the discrete form to give an approximate solution to Eq.~\eqref{FKeq} as follows
\begin{equation}\label{BreatherSoln}
    u_n^{}(t) = 4\arctan\frac{\eta\cos[\zeta\omega_{\rm DB}^{}(t - v_{\rm DB}^{}n)]}{\omega_{\rm DB}^{}\cosh[\zeta\eta(n - v_{\rm DB}^{}t)]},
\end{equation}
where
\begin{equation}
    \eta = \sqrt{1 - \omega^{2}_{\rm DB}},\hspace{0.5cm}\zeta = \frac{1}{\sqrt{1 - v^{2}_{\rm DB}}}.
\end{equation}
Amplitude of standing DB ($v_{\rm DB}^{}=0$) is
\begin{equation}\label{BreatherSoln1}
    A_{\rm DB}^{} = 4\arctan\frac{\eta}{\omega_{\rm DB}^{}}.
\end{equation}

The red solid line in Fig.~\ref{fig:fig10}(b) shows the relation between the DB frequency $f_{\rm DB}^{}=\omega_{\rm DB}^{}/2\pi$ and the DB amplitude $A_{\rm DB}^{}$ for soft-type anharmonicity obtained from the analytical solution \eqref{BreatherSoln1}. Blue dashed line gives the frequency-amplitude relation obtained with the machine accuracy for the on-site DB~\cite{Machine}. Accuracy of the analytical solution \eqref{BreatherSoln1} is much better than in the case of the hard-type of anharmonicity, and, in contrast to that case, here the accuracy increases with decreasing amplitude. Numerical results for the chaotic DBs (scattered dots) obtained from the simulation of the modulational instability of the $\Gamma$-point mode with the amplitude $A=0.6$ are also shown. The analytical solution \eqref{BreatherSoln1} is in a good agreement with the simulation results (within 1.5\%).

\section{Conclusions}
\label{Conclusion}

In the present study, the effect of DBs on the total to kinetic energy ratio of the chain is discussed and related to the specific heat of a nonlinear chain. Chaotic DBs arise in the chain as a result of the modulational instability of particular extended vibrational modes, namely, the zone-boundary mode ($q=\pi$) for the case of hard-type anharmonicity and the $\Gamma$-point mode ($q=0$) for the case of soft-type anharmonicity. The analytical results presented in Sec.~\ref{DB} prove that the localized modes observed in the system are DBs. 

As it can be seen in Fig.~\ref{fig5}, when DBs are excited in the chain with the hard-type anharmonicity, the specific heat is about 10\% lower as compared to thermal equilibrium. For the soft-type anharmonicity (see Fig.~\ref{fig9}), the specific heat reduces by about 2\% during the transition from the regime with DBs to thermal equilibrium. This means that DBs reduce (increase) the heat capacity of the nonlinear chain with hard-type (soft-type) anharmonicity. 

The results obtained here have a very simple physical interpretation. DBs in the hard-type anharmonicity chain have frequencies above the phonon spectrum, so that their excitation would increase the velocities of particles and hence the kinetic energy (or temperature) of the system. In this situation, DBs are responsible for a decrease of heat capacity because temperature or kinetic energy is in the denominator of Eq.~(\ref{HeatCap}) or Eq.~(\ref{HeatCapHere}). For the chain with the soft-type anharmonicity the effect of DBs on heat capacity is opposite because DBs have frequencies below the phonon spectrum and their appearance in the system will lead to a decrease in particle velocities and kinetic energy (or temperature) of the system.

In this study, chaotic DBs emerged as a result of instability of the zone-boundary mode were analyzed. No doubt that the conclusions made are valid for thermally populated DBs in the regime of thermal equilibrium as well. On the other hand, the effect will be considerably weaker, since the portion of energy carried by DBs in thermal equilibrium is much smaller than in the far-from-equilibrium case considered here.

Having this in mind, one can deduce that the heat capacity of the crystals having a gap in the phonon spectrum and supporting soft-type anharmonicity DBs (e.g., NaI~\cite{NaI1,NaI2,NaIDBMD1,NaIDBMD2,NaIDBMD3}, ordered alloys~\cite{OA0,OA1,OA2,OA3}, and graphane~\cite{AbInitgraphane}) should increase due to the excitation of DBs. For the crystals without a gap in the phonon spectrum (e.g., pure metals~\cite{M1,M2,M3,M4,M5,M6} and covalent crystals~\cite{SiGe,Diamond}) only hard-type anharmonicity DBs can exist and their excitation will reduce heat capacity. 

Note that in the experimental work~\cite{UheatCapac} an increase of heat capacity of alpha-uranium at high temperatures was related to the contribution from DBs. Apparently this is a misleading interpretation since DBs in alpha-uranium are of the hard-type anharmonicity~\cite{M4} and they can only {\em reduce} the heat capacity.

In future studies, it is planned to analyze the effect of DBs on other macroscopic properties of nonlinear chains and crystal lattices, e.g., on the elastic constants and thermal expansion. These properties could not be analyzed in the present study since the model with the on-site potential was considered. The on-site potential precludes the free thermal expansion of the chain and alters the mechanical response of the chain to external loads. On the other hand, models without on-site potential support only hard-type anharmonicity DBs and they cannot be used to study the effect of the type of anharmonicity, which was addressed here.  

Overall, the works devoted to the effect of DBs on macroscopic properties of nonlinear lattices will suggest the ways of indirect detection of DBs in crystals by measuring macroscopic properties sensitive to the presence of DBs.

\section*{Acknowledgments}

The work of E.A.K.\ was supported by the Grant of the President of the Russian Federation for State support of young Russian scientists (No.\ MD-3639.2019.2). The work of D.X.\ is supported by NNSF (Grant No.~11575046) of China, NSF (Grant No.~2017J06002) of Fujian Province of China. The work of V.A.G.\ was supported by the MEPhI Academic Excellence Project. S.V.D.~acknowledges the support of the Russian Foundation for Basic Research, Grant No. 19-02-00971. The work was partly supported by the State assignment of IMSP RAS No. AAAA-A17-117041310220-8. The work of O.B.N.\ was supported by the State Assignement of ICMM UB RAS, Government Contract No.\ AAAA-A19-119013090021-5.

\end{document}